\documentclass[twocolumn,english,showpacs,amsmath,amssymb,prl,superscriptaddress,floatfix]{revtex4}
\usepackage[T1]{fontenc}
\usepackage[latin9]{inputenc}
\usepackage{graphicx}
\usepackage{amssymb}

\usepackage{babel}
\usepackage{color}

\begin{document}

\title{Mesoscopic Rydberg Gate based on Electromagnetically Induced Transparency}

\author{M. M\"uller}

\affiliation{Institute for Theoretical Physics, University of Innsbruck, and Institute
for Quantum Optics and Quantum Information of the Austrian Academy
of Sciences, Innsbruck, Austria}

\author{I. Lesanovsky}

\affiliation{Institute for Theoretical Physics, University of Innsbruck, and Institute
for Quantum Optics and Quantum Information of the Austrian Academy
of Sciences, Innsbruck, Austria}

\author{H. Weimer}

\affiliation{Institute for Theoretical Physics III, University of Stuttgart, Pfaffenwaldring
57, 70550 Stuttgart, Germany}

\author{H.P. B\"uchler}

\affiliation{Institute for Theoretical Physics III, University of Stuttgart, Pfaffenwaldring
57, 70550 Stuttgart, Germany}

\author{P. Zoller}

\affiliation{Institute for Theoretical Physics, University of Innsbruck, and Institute
for Quantum Optics and Quantum Information of the Austrian Academy
of Sciences, Innsbruck, Austria}

\date{\today}
\begin{abstract}
\label{txt:abstract} We demonstrate theoretically a parallelized
\textsc{c-not} gate which allows to entangle a mesoscopic ensemble
of atoms with a single control atom in a single step, with high fidelity
and on a microsecond timescale. Our scheme relies on the strong and
long-ranged interaction between Rydberg atoms triggering Electromagnetically
Induced Transparency (EIT). By this we can robustly implement a conditional
transfer of all ensemble atoms between two logical states, depending
on the state of the control atom. We outline a many body interferometer
which allows a comparison of two many-body quantum states by performing
a measurement of the control atom.
\end{abstract}


\maketitle
Atoms excited by laser light to high-lying Rydberg states interact
via strong and long-range dipole-dipole or Van der Waals forces \cite{Gallagher}.
Level shifts associated with these interactions can be used to block
transitions of more than one Rydberg excitation in mesoscopic atomic
ensembles. This ``dipole blockade'' \cite{JakschLukinMolmer} mechanism underlies the formation of ``superatoms'' in atomic gases with a single Rydberg excitation
shared by many atoms within a blockade radius. Furthermore, this provides
the basis for fast two-qubit gates between pairs
of atoms in optical or magnetic trap arrays. Recently, these superatoms
and Rydberg gates have been demonstrated in the laboratory by several
groups in remarkable experiments \cite{RydExperiments,TwoTraps}, also combining the tools of Electromagnetically
Induced Transparency (EIT) and Rydberg blockade \cite{Adams}.  Building
on these achievements, a future challenge is to develop and extend
Rydberg-based protocols towards single step \emph{many atom entanglement}.
Here we propose and analyze a fast high-fidelity many-particle gate by combining elements of EIT and Rydberg interactions, which entangles \textit{in a single step} a control atom with a mesoscopic number of atoms $N$. As discussed below, such a mesoscopic parallel Rydberg gate has immediate applications in quantum information processing and entanglement-based many particle interferometry, and represents a quantum amplifier or single atom transistor \cite{Micheli}.
\begin{figure}[htb]

\begin{centering}
\includegraphics[width=8.5cm]{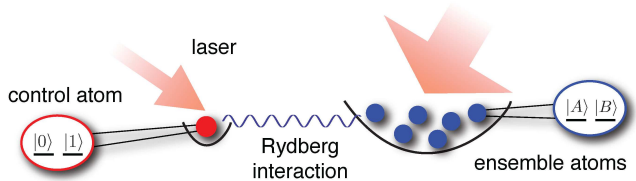}
\par\end{centering}

\caption{In the envisioned setup the quantum state of an atomic ensemble is
manipulated depending on the state of a single control atom. The atomic
ensemble can consist of atoms in a single trap or of atoms being confined
in a lattice.}

\label{fig:setup}
\end{figure}

We envision a setup as illustrated in Fig.~1. A control atom and a
mesoscopic ensemble of atoms are stored in two separate trapping potentials,
e.g. in two dipole traps as in Ref. \cite{TwoTraps}, or in
large-spacing optical lattices or magnetic trap arrays
\cite{NelsonWhitlock}. Our goal is the implementation of the operation
\textsc{c-not}$^{N}$, defined by \begin{eqnarray}
|0\rangle|A^{N}\rangle & \rightarrow & |0\rangle|A^{N}\rangle,\quad\,|0\rangle|B^{N}\rangle\rightarrow|0\rangle|B^{N}\rangle,\label{eq:trafo}\\
|1\rangle|A^{N}\rangle & \rightarrow & |1\rangle|B^{N}\rangle,\quad\,|1\rangle|B^{N}\rangle\rightarrow|1\rangle|A^{N}\rangle,\nonumber \end{eqnarray}
where $|0\rangle$, $|1\rangle$ and $|A\rangle$ and $|B\rangle$
denote long-lived ground states of the control and ensemble atoms,
respectively. The gate consists of a conditional swap of the two internal
states of $N$ ensemble atoms,
where we have adopted the notation $|A^{N}\rangle\equiv\bigotimes_{k=1}^{N}|A\rangle_{k}$
and $|B^{N}\rangle\equiv\bigotimes_{k=1}^{N}|B\rangle_{k}$. The gate
(\ref{eq:trafo}) corresponds to a Schr\"odinger-cat or GHZ-type beam
splitter: $\left(\alpha|0\rangle+\beta|1\rangle\right)|A^{N}\rangle\rightarrow\alpha|0\rangle|A^{N}\rangle+\beta|1\rangle|B^{N}\rangle$. The resulting state constitutes an important resource for quantum
computing, and provides a basic ingredient for Heisenberg limited
interferometry \cite{Raimond06}.

The basic elements and steps in our realization of the gate (\ref{eq:trafo})
are: (i) the control atom can be individually addressed
and laser excited to a Rydberg state conditional to its internal state,
thus (ii) turning on or off the strong long-range Rydberg-Rydberg
interactions of the control with ensemble atoms, which (iii) via EIT-type
interference suppresses or allows the transfer of all ensemble
atoms from $|A\rangle$ or $|B\rangle$ conditional to the state of
the control atom. Among the distinguishing features of our protocol
is high fidelity for moderately sized atomic ensembles spread
out over several micrometers. It does not require individual 
addressing of the ensemble atoms, in contrast to a possible implementation of the gate (\ref{eq:trafo}) by a sequence of $N$ two qubit gates. It is robust with respect
to inhomogeneous interparticle distances and varying interaction strengths
and can be carried out on a microsecond timescale. Furthermore, we find that mechanical
effects caused by strong forces between Rydberg atoms will not
spoil the fidelity of the gate operation.

\begin{figure}[htb]

\begin{centering}
\includegraphics[width=8.5cm]{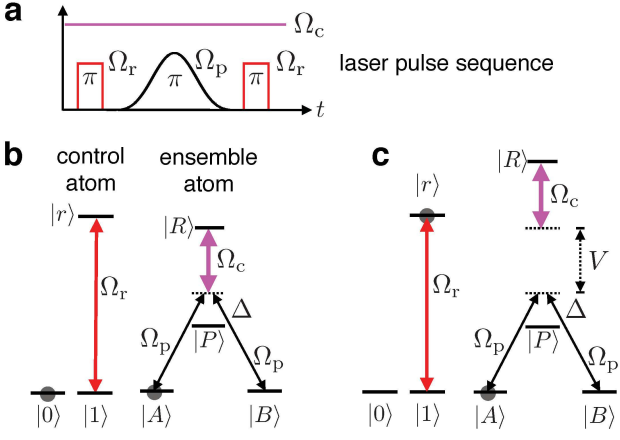}
\par\end{centering}

\caption{ \textbf{a:} Sequence of laser pulses (not to scale). \textbf{b:}
Electronic level structure of the control and ensemble atoms. The
ground state $\left|1\right\rangle $ is resonantly coupled to the
Rydberg state $\left|r\right\rangle $. The states $\left|A\right\rangle $
and $\left|B\right\rangle $ are off-resonantly coupled (detuning
$\Delta$, Rabi frequency $\Omega_{\mathrm{p}}$) to $\left|P\right\rangle $.
A strong laser with Rabi frequency $\Omega_{\mathrm{c}}\gg\Omega_{\mathrm{p}}$
couples the Rydberg level $\left|R\right\rangle $ to $\left|P\right\rangle $
such that $\left|R\right\rangle $ is in two-photon resonance with
$\left|A\right\rangle $ and $\left|B\right\rangle $. In this situation
(known as EIT) Raman transfer from $\left|A\right\rangle $ to $\left|B\right\rangle $
is inhibited. \textbf{c:} With the control atom excited to $\left|r\right\rangle $
the two-photon resonance condition is lifted as the level $\left|R\right\rangle $
is shifted due to the interaction energy $V$ between the Rydberg
states, thereby enabling off-resonant Raman transfer from $\left|A\right\rangle $
to $\left|B\right\rangle $.}

\label{fig:level_structure}
\end{figure}

Let us now discuss the concrete physical implementation of the gate
(\ref{eq:trafo}) and introduce intermediate states for
the control and ensemble atoms (see Fig.~\ref{fig:level_structure}).
For the control atom we consider the Rydberg level $|r\rangle$ which is resonantly
coupled to $|1\rangle$ by a laser with (two-photon)-Rabi
frequency $\Omega_{\mathrm{r}}$. In the rotating wave approximation
the corresponding Hamiltonian reads $H_{\mathrm{r}}=\left[(\hbar\Omega_{\mathrm{r}}/2)\left|1\right\rangle \left<r\right|+\mathrm{h.c.}\right]$.
The ensemble atoms possess the two stable ground states $|A\rangle$
and $|B\rangle$, the intermediate state $|P\rangle$ and a Rydberg
state $|R\rangle$. The state $|P\rangle$ can be a p-state of an
alkali metal atom, e.g.~$5^{2}P_{3/2}$ in case of $^{87}$Rb, and
possesses a lifetime $\gamma_{p}^{-1}$ of tens of nanoseconds. The
ground states are off-resonantly coupled (detuning $\Delta$ with $\Delta\gg\gamma_{p}$)
to $|P\rangle$ by two Raman lasers, which for simplicity are assumed to have the same Rabi frequency
$\Omega_{\mathrm{p}}$ (see Fig.~\ref{fig:level_structure}b). A second
laser with Rabi frequency $\Omega_{\mathrm{c}}$ ($\Delta \gg \Omega_{c} > \Omega_{p}$) couples $|R\rangle$ and $|P\rangle$ such that the two ground states are
in two photon resonance with $|R\rangle$. 

We now outline the conditional transfer of the
ensemble atoms from the state $\left|A^{N}\right\rangle $ to $\left|B^{N}\right\rangle$.
We start with the case of non-interacting ensemble atoms since
it can be treated in a single particle picture. Subsequently,
we discuss the effects caused by the interaction among
the ensemble atoms. We distinguish two
cases: For the control atom in $|0\rangle$ we intend to block the transfer
$\left|A^{N}\right\rangle \rightarrow\left|B^{N}\right\rangle $, whereas for the initial state $|1\rangle$ the transfer shall be enabled. In both cases the
same sequence of three laser pulses, as sketched in Fig.~\ref{fig:level_structure}a,
is applied: a short $\pi$-pulse on the control atom,
a smooth Raman $\pi$-pulse $\Omega_{\mathrm{p}}(t)$ with $\int_{0}^{T}\mathrm{d}t\,\Omega_{\mathrm{p}}^{2}(t)/(2\Delta)=\pi$
acting on all ensemble atoms, and a second $\pi$-pulse on
the control atom.

\begin{figure}[htb]

\begin{centering}
\includegraphics[width=8.5cm]{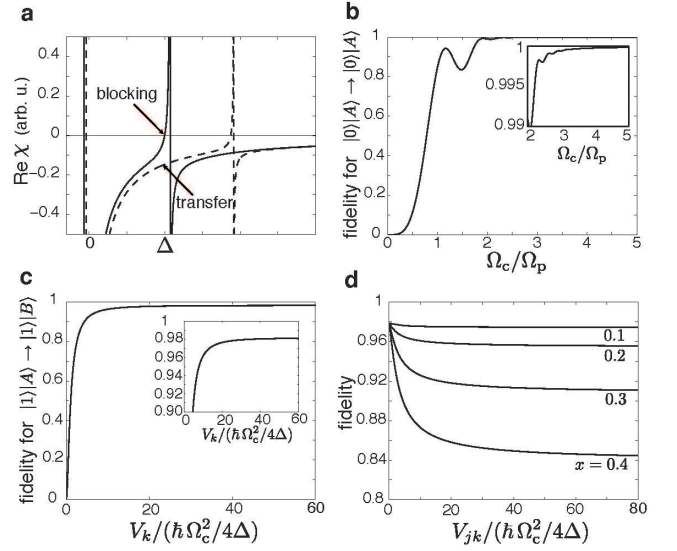}
\par\end{centering}
\caption{\textbf{a:} Linear susceptibility (not to scale) with respect to the Raman laser as a function of its detuning $\delta$
from the $\left|P\right>$-level for blocked
transfer (solid curve) and in the unblocked case (dashed curve). \textbf{b:}
Efficiency of the blocking (I) as a function of $\Omega_{\mathrm{c}}/\Omega_{\mathrm{p}}$.
For $\Omega_{\mathrm{c}}/\Omega_{\mathrm{p}}>2$ the transfer of the
ensemble atoms from $\left|A\right\rangle $ to $\left|B\right\rangle $
is blocked with more than $99\,\%$ fidelity. \textbf{c:} Transfer efficiency
in the unblocked case (II) as a function
of the interaction between the control and one ensemble atom
($\Omega_{\mathrm{c}}=6\,\Omega_{p}$). \textbf{d:} Fidelity of the process $1/\sqrt{2}\left(\left|0\right\rangle +\left|1\right\rangle \right)\left|AAA\right\rangle \rightarrow1/\sqrt{2}\left(\left|0\right\rangle \left|AAA\right\rangle +\left|1\right\rangle \left|BBB\right\rangle \right)$
for three ensemble atoms as a function of their interaction strength
$V_{jk}$ and the ratio $x_{\mathrm{max}}=\sqrt{2}\mathrm{max}(\Omega_{\mathrm{p}})/\Omega_{\mathrm{c}}$. We chose the worst case scenario, i.e. all $V_{jk}$ are equal.
The interaction between control atom and ensemble atoms was $V_{k}=10\,\hbar\Omega_{\mathrm{c}}^{2}/\Delta$
for all four curves, giving rise to a maximal allowed distance between
the control atom and the ensemble atoms of 2.2 $\mu$m (1.4 $\mu$m)
for the ratio $x=0.4$ ($x=0.1$). We have chosen $\mathrm{max}(\Omega_{\mathrm{p}})=2\pi\times70$
MHz, $\Delta=2\pi\times1.2$ GHz in \textbf{b}-\textbf{d}, and atomic
parameters of $^{87}$Rb (see text).}
\label{fig:eit_fidelity}
\end{figure}

\textit{(I) - blocking:} $\left|0\right\rangle \left|A^{N}\right\rangle \rightarrow\left|0\right\rangle \left|A^{N}\right\rangle $
- The blocking  can be conveniently understood in terms of adiabatic passage along dark states of an effective Hamiltonian for the $k$-th ensemble atom,
\begin{eqnarray}\label{eff}
H_{k}/\epsilon = x^2 \left|+\right>_k\!\left<+\right|\:
+\: \left|R\right>_k\!\left<R\right| \: +\: x
\Big(\left|+\right>_k\!\left<R\right| + \mathrm{h.c.}\Big),
\end{eqnarray}
obtained by adiabatically eliminating
the far-detuned  $\left|P\right\rangle $ state from the four-level system depicted in Fig.~\ref{fig:level_structure}b in the limit $\Delta\gg\Omega_{\mathrm{c}},\Omega_{\mathrm{p}}$.  
In Eq.~(\ref{eff}) we have defined a characteristic energy scale $\epsilon=\hbar\Omega_{\mathrm{c}}^{2}/(4\Delta)$,
the states $\left|\pm\right\rangle =(1/\sqrt{2})[\left|A\right\rangle \pm\left|B\right\rangle ]$
and the rescaled, dimensionless Raman laser Rabi frequency $x(t)=\sqrt{2}\Omega_{\mathrm{p}}(t)/\Omega_{\mathrm{c}}$.
%
We are interested in the regime $x\ll1$, in which $H_{k}$ describes
the EIT scenario \cite{EIT}. The solid curve in Fig.~\ref{fig:eit_fidelity}a
shows the susceptibility $\chi(\delta)$ with respect to $\Omega_{\mathrm{p}}$ as a function of the detuning $\delta$ of the Raman lasers from the $\left|P\right>$-state.
We work on two photon-resonance $\delta=\Delta$ with $\chi(\Delta)=0$. Here the ensemble atoms become 'transparent' for the Raman lasers which then do not couple the states $\left|A\right\rangle$ and $\left|B\right\rangle$ anymore. In this case $H_{k}$ has two dark states
\begin{eqnarray}
\left|d_{1}\right\rangle _{k}=\left|-\right\rangle ,\quad\left|d_{2}\right\rangle _{k}=(1+x^{2})^{-1/2}[\left|+\right\rangle _{k}-x\left|R\right\rangle _{k}].\end{eqnarray}
For the control atom initially in $\left|0\right\rangle $
the first $\pi$-pulse has no effect. During the smooth Raman pulse the $k$-th ensemble atom will adiabatically
follow the dark state $\left|d\right\rangle _{k}=(1/\sqrt{2})\left[\left|d_{1}\right\rangle _{k}+\left|d_{2}\right\rangle _{k}\right]$,
thereby starting and ending in $\left|A\right\rangle _{k}$.
The remaining (non-dark) states are separated by an energy of at least
$\epsilon$. This strongly inhibits non-adiabatic losses by Landau-Zener
transitions, which limit the blocking fidelity and
occur with a small probability $\propto x^{6}$. Non-adiabatic couplings
to the other dark state are absent. From Fig.~\ref{fig:eit_fidelity}b
we see that the transfer is blocked with more than $99\,\%$ fidelity \footnote{We define the fidelity by the overlap of the desired and the obtained final state, $F=|\left<\mathrm{desired}\mid\mathrm{obtained}\right>|^2$.}
if $\Omega_{\mathrm{c}}/\mathrm{max}(\Omega_{\mathrm{p}}(t))>2$.
After the second (ineffective) $\pi$-pulse on the control atom we
have performed the step $\left|0\right\rangle \left|A^{N}\right\rangle \rightarrow\left|0\right\rangle \left|A^{N}\right\rangle $.

\textit{(II) - transfer:} $\left|1\right\rangle \left|A^{N}\right\rangle \rightarrow\left|1\right\rangle \left|B^{N}\right\rangle $
- If the control atom is initially in $\left|1\right\rangle $
the first $\pi$-pulse transfers it to $\left|r\right\rangle $. Since the control and the ensemble atoms
interact via $H_{\mathrm{ce}}=\sum_{k}V_{k}\left|r\right\rangle \left<r\right|\otimes\left|R\right\rangle _{k}\left<R\right|$
the Rydberg level of the $k$-th ensemble atom is now shifted by the
energy $V_{k}>0$ (see Fig.~\ref{fig:level_structure}c). This interaction-induced
energy shift lifts the two-photon resonance condition, which is crucial
to block the Raman transfer from $\left|A\right\rangle _{k}$ to $\left|B\right\rangle _{k}$.
Now, the Raman laser beams no longer address the point of vanishing
susceptibility $\chi(\Delta)=0$ (cf.~dashed curve in Fig.~\ref{fig:eit_fidelity}a),
but couple off-resonantly to $\left|P\right\rangle _{k}$ and
thereby realize the transfer from $\left|A\right\rangle _{k}$
to $\left|B\right\rangle _{k}$. In Fig.~\ref{fig:eit_fidelity}c
the efficiency of the transfer $\left|1\right\rangle \left|A\right\rangle _{k}\rightarrow\left|1\right\rangle \left|B\right\rangle _{k}$
is shown as a function of $V_{k}$. Theoretically, ideal transfer
is achieved for $V_{k}\gg\epsilon$ but even for $V_{k}>40\epsilon$
the fidelity exceeds $98\,\%$.

The upper limit for the transfer fidelity is set by three factors.
First, radiative decay from the p-state occurs with
a probability $\sim\gamma_{p}/\Delta\ll1$ during the Raman transfer.
Once the transfer has taken place the control atom is returned to
$\left|1\right\rangle $ through the second $\pi$-pulse,
and eventually the step $\left|1\right\rangle \left|A^{N}\right\rangle \rightarrow\left|1\right\rangle \left|B^{N}\right\rangle $
is completed \footnote{More precisely, the step $\left|1\right>\left|A^N\right> \rightarrow -(-1)^N \left|1\right>\left|B^N\right>$ is realized. These additional phase factors can be avoided by controlling the relative phases of the laser pulses.}. Second, during the Raman pulse the control atom
resides in $\left|r\right\rangle $ (lifetime $\tau_{\mathrm{r}}$)
for a time $T$. In order to minimize radiative decay
from $\left|r\right\rangle $, which reduces the transfer fidelity by a factor $\exp(-T/\tau_{\mathrm{r}})$
(independently of $N$), the Raman pulse has to be carried out much
faster than $\tau_{\mathrm{r}}$, i.e. $T<1\mu$s. Third, mechanical forces can occur if the control atom \emph{and} an ensemble atom reside in a Rydberg
state at a time. This would cause entanglement of internal and external
degrees of freedom. However, since the probability for a double occupation
of the Rydberg state is $\propto x^{2}(\epsilon/V_{k})^{2}$ the corresponding
loss of fidelity is negligibly small.

As the procedure is time-reversal symmetric, the inverse operation $\left|0\right\rangle \left|A^{N}\right\rangle \rightarrow\left|0\right\rangle \left|A^{N}\right\rangle $
and $\left|1\right\rangle \left|B^{N}\right\rangle \rightarrow\left|1\right\rangle \left|A^{N}\right\rangle $
is achieved by precisely the same pulse sequence.

Let us now extend the discussion to many interacting ensemble atoms.
Ensemble-ensemble interactions $H_{\mathrm{ee}}=\sum_{k>j}V_{jk}\left|R\right\rangle_{j}\!\left<R\right|\otimes\left|R\right\rangle_{k}\!\left<R\right|$
are of no consequence for transfer step \textit{(II)} provided
$V_{jk}\geq0$, which can be ensured by the proper choice of the Rydberg
state $\left|R\right\rangle $. Since in this step the Rydberg level
is anyway shifted by $H_{\mathrm{ce}}$ a further shift by $H_{\mathrm{ee}}$
will have no effect. However, the influence of $H_{\mathrm{ee}}$
on step \textit{(I)} is more delicate as the blocking crucially relies on the EIT condition. Fig.~\ref{fig:eit_fidelity}d shows the fidelity for generating the state $|0\rangle|A^{N}\rangle+|1\rangle|B^{N}\rangle$
for three ensemble atoms as a function of their mutual interaction
and $x_{\mathrm{max}}=\sqrt{2}\mathrm{max}\left(\Omega_{\mathrm{p}}(t)\right)/\Omega_{c}$.
The fidelity decreases with increasing ensemble-ensemble interaction. Surprisingly, however, it quickly approaches
a constant value as $V_{jk}$ is further increased. This asymptotic
value increases the smaller the parameter $x_{\mathrm{max}}$. We
will now show that in the limit $x_{\mathrm{max}}\ll1$ the blocking
works with high fidelity, \emph{independently} of the interaction strength among the ensemble atoms. Consequently,
in this limit the maximally achievable fidelity of the gate
(\ref{eq:trafo}) becomes independent of $H_{\mathrm{ee}}$ and is
solely determined by the imperfections discussed in \textit{(I)} and
\textit{(II)}.

The initial state of the ensemble atoms can be written as a sum of direct products of
the single particle dark states, which for two atoms takes the form
$\left|AA(0)\right\rangle =(1/2)\left[\left|d_{1}d_{1}\right\rangle
+\left|d_{1}d_{2}\right\rangle +\left|d_{2}d_{1}\right\rangle
+\left|d_{2}d_{2}\right\rangle \right]$.  While $\left|d_{1}d_{1}\right\rangle $, $\left|d_{1}d_{2}\right\rangle $ and
$\left|d_{2}d_{1}\right\rangle $ remain exact dark states for finite
$x(t)\neq0$, the state $|d_{2} d_{2}\rangle$ contains a fraction of the two atom
Rydberg state $|R R\rangle $, and due to the Rydberg interaction evolves into a new state $|g\rangle$ under the adiabatic time evolution $x(t)$. This new state
acquires an energy shift $E_{g}$, causing a dynamical phase shift which is the dominant mechanism for the reduction of the blocking fidelity. For weak interactions
$V_{jk}\ll\epsilon$ perturbation theory yields
$E_{g}(t)=x^4(t) V_{12}$, while in for strong interactions $V_{j k} \gg
\epsilon$ it reaches the asymptotic behavior $E_{g}(t)\approx2\epsilon
x^{4}(t)(1-2\, V_{12}^{-1})$. Consequently, the acquired phase shift is bounded
by the worst case scenario of strong interactions with $V_{j k} \gg \epsilon$.
Then, the 'grey' state $|g\rangle$ in the
present situation with $\Omega_{c} \ll \Delta$ reduces to the collective state
\begin{eqnarray}
\left|g\right\rangle =(1+x^{4})^{-1/2}\left[(1-x^{2})\left|++\right\rangle
-x(\left|+R\right\rangle +\left|R+\right\rangle )\right],\end{eqnarray}
which has the energy $E_{g}(t)=2\epsilon x^{4}(t)$, and contains an admixture of
the 'superatom' state $\frac{1}{\sqrt{2}}(\left|+R\right\rangle
+\left|R+\right\rangle )$ when the Raman lasers are on. The acquired dynamical phase is
$2\phi$ with $\phi=\int_{0}^{T}\mathrm{d}tE_{g}(t)/\hbar\approx(35/96)2\pi
x_{\mathrm{max}}^{2}$
\footnote{The prefactor stems from the particular choice of the Raman
laser profile: $\Omega_{p}(t)=\sqrt{(16\pi\Delta)/(3T)}\sin^{2}(\pi
t/T)$.}.
After the Raman pulse we find $\left|AA(T)\right\rangle
=(1/2)\left[\left|d_{1}d_{1}\right\rangle +\left|d_{1}d_{2}\right\rangle
+\left|d_{2}d_{1}\right\rangle +e^{-2i\phi}\left|g\right\rangle \right]$, giving
rise to a blocking fidelity $F_{b}=\left|\left<AA(0)\mid
AA(T)\right\rangle \right|^{2}=(1/16)\left|3+e^{-2i\phi}\right|^{2}$. This analysis can be generalized for $N$ ensemble atoms where one finds $N+1$ 'true' dark states and $2^{N}-(N+1)$ 'grey' states, which sustain
energy shifts of at most $\epsilon\, N(N-1)x^{4}(t)$ during the Raman pulse.
The fidelity is then $F_{b}=\left|\sum_{m=0}^{N} N!/(m!(N-m)!2^N)\exp(-im(m-1)\phi)\right|^{2}$
with $\phi\ll1$ as defined above. 
Note that the fidelity can
be improved by suppressing the ensemble interaction with a suitable choice of
the Rydberg state and/or a cancelation of the leading interaction by the
combination of static and microwave fields as was recently proposed for polar
molecules \cite{Buechler07}. For strong interactions the probability of finding two ensemble atoms in the Rydberg state is of the order
$N^{2}x^{4}(\mathrm{min}(V_{jk})/\epsilon)^{-2}$,
such that mechanical effects, which might reduce the fidelity, are
negligible.

The numbers presented in this work (Fig.~\ref{fig:eit_fidelity})
have been calculated for $^{87}$Rb. The Rydberg states of the control
and the ensemble atoms are $50s$ and $49s$, respectively \footnote{No particular form of the interaction potential is needed; accidental resonances should be avoided (see, e.g., T. G. Walker, M. Saffman, Phys. Rev. A 77, 032723 (2008)).}. The $C_{6}$ coefficients of the corresponding Van der Waals interaction have been taken from \cite{Singer05}. The lifetime of the control atom is $\tau_{c}=66\,\mu\mathrm{s}$.  The detuning of the Raman lasers is $\triangle=2\pi\times1.2\,\mathrm{GHz}$, the duration $T=0.44\,\mu\mathrm{s}$ and the decay rate of the p-state $\gamma_{p}=36\,\mathrm{MHz}$. Larger interaction energies and correspondingly larger distances between the control and the ensemble atoms can be reached
by choosing Rydberg states of higher principal quantum number and/or
working with permanent induced dipole moments.

\begin{figure}[htb]
\begin{centering}
\includegraphics[width=8.5cm]{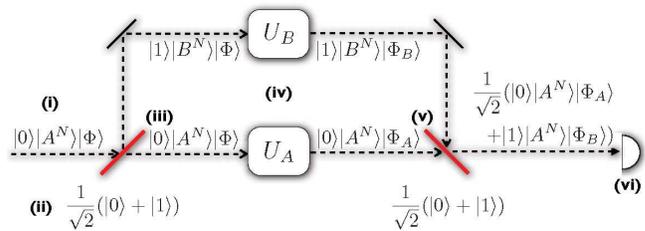}
\par\end{centering}

\caption{The gate (\ref{eq:trafo}) is the fundamental building
block of a many-particle interferometer where the overlap of two many-particle
wave functions can be measured. (i) Initial state preparation. (ii) The control atom is prepared in $(1/\sqrt{2})(\left|0\right>+\left|1\right>)$. (iii) Gate (Eq.~(\ref{eq:trafo})). (iv) Internal-state-dependent evolution, governed by $U_A$ and $U_B$. (v) Recombination of the interferometer arms by (\ref{eq:trafo}). (vi) Measurement of the control atom in the basis $\left|c_\pm\right>=(1/\sqrt{2})(\left|0\right>\pm\left|1\right>)$ yields access to the wave function overlap $\left<\Phi_{A}\mid\Phi_{B}\right>=\left<\Phi\right|U_{A}^{\dagger}U_{B}\left|\Phi\right>$.}
\label{fig:interferometer}
\end{figure}

Finally, we briefly comment on the novel possibilities offered by
our mesoscopic Rydberg gate in the context of quantum dynamics of
atomic condensed matter physics. In Fig.~1 the atomic ensemble can
represent cold atoms in an optical lattice as a quantum simulator for
a Hubbard model. Since optical lattices can be state dependent, atoms
in $|A\rangle$, and $|B\rangle$ can be governed by Hubbard Hamiltonians
with different (time-dependent) parameters, generating for an initial
quantum state or phase $|\Phi\rangle$ in the lattice a different
time evolution $U_{A,B}|\Phi\rangle\rightarrow|\Phi_{A,B}\rangle$.
The gate (\ref{eq:trafo}) allows the preparation of such a mesoscopic
superposition of quantum phases on time scales fast compared with
the lattice dynamics, entangled with the control atom, $\left|0\right\rangle \left|A^{N}\right\rangle \left|\Phi_{A}\right\rangle +\left|1\right\rangle \left|B^{N}\right\rangle \left|\Phi_{B}\right\rangle $.
A many particle interferometer, as described in Fig.~\ref{fig:interferometer},
will provide via measurement of the control atom the overlap $\langle\Phi_{A}|\Phi_{B}\rangle$ \cite{Zanardi07}.
This compares \emph{many body quantum states and their dynamics} on
the level of the full wave function, at least of a mesoscopic scale,
to be compared with low order correlation functions accessed in traditional
condensed matter and cold atom experiments.

M.M. thanks A. Browaeys and P. Grangier for discussions. Support by the Institute for Quantum Optics and Quantum Information, the FWF through SFB FOQUS, and the EU projects SCALA and NAMEQUAM as well as by the DFG within SFB/TRR21 is acknowledged.

\end{document}